\documentclass[12pt]{iopart}
\usepackage{graphicx}

\begin{document}

\title[Characterization of single photon sources for radiometry applications]{Characterization of single photon sources for radiometry applications at room temperature}

\author{Hee-Jin Lim$^1$,  Dong-Hoon Lee$^1$, In-ho Bae$^1$,
	Kwang-Yong Jeong$^2$, 
	Christoph Becher$^3$,  
	Sejeong Kim$^4$,
	Igor Aharonovich$^{4,5}$,
	and Kee Suk Hong$^{1,*}$}

\address{$^1$ Korea Research Institute of Standards and Science (KRISS), Daejeon 34113, South Korea}
\address{$^2$ Department of Physics, Korea University, Seoul 02841, South Korea}
\address{$^3$ Naturwissenschaftlich-Technische Fakult\"{a}t, Fachbereich Physik, Universit\"{a}t des Saarlandes, Campus E 2.6, 66123 Saarbr\"{u}cken, Germany}
\address{$^4$ School of Mathematical and Physical Sciences, University of Technology Sydney, Ultimo, NSW, 2007, Australia}
\address{$^5$ ARC Centre of Excellence for Transformative Meta-Optical Systems, University of Technology Sydney, Ultimo, NSW, 2007, Australia}

\ead{hongi2011@kriss.re.kr}
\vspace{10pt}

\begin{abstract}
A single photon source with high repeatability and low uncertainties is the key element for few-photon metrology based on photon numbers.
While low photon number fluctuations and high repeatability are important figures for qualification as a standard light source, these characteristics are limited in single photon emitters by some malicious phenomena like blinking or internal relaxations to varying degrees in different materials.
This study seeks to characterize photon number fluctuations and repeatability for radiometry applications at room temperature. 
For generality in this study, we collected photon statistics data with various single photon emitters of $g^{(2)}(0) < 1$ at low excitation power and room temperature in three material platforms: silicon vacancy in diamond, defects in GaN, and vacancy in hBN.
We found common factors related with the relaxation times of the internal states that indirectly affect photon number stability. We observed a high stability of photon number with defects in GaN due to faster relaxations compared with vacancies in hBN, which on the other hand produced high rates ($> 10^6$) of photons per second.
Finally, we demonstrate repeatable radiant flux measurements of a bright hBN single photon emitter for a wide radiant flux range from a few tens of femtowatts to one picowatt.
\end{abstract}

%
\vspace{2pc}

\submitto{Materials for Quantum Technology}

\maketitle
%
%
\section{Introduction}
Advances in photon measurements have stimulated innovation in encrypted communications and information processing, distinguished by a single datum being encoded in a single photon \cite{Tittel:01,chen:06}.
As relevant techniques mature, accuracy in photon generation and detection is growing to compete with the present standard of radiative flux.
A new standard is thus being defined by photon flux, in which counting is a central element for information processing with photons, and which is differentiated from the present standard where absoluteness is given by foreign physical systems based on electrical current and temperature \cite{Willson:73,Martin_1985,QCandelta:07,Zwinkels:10}.

For the realization of this new standard, the accurate calibration of photon flux as well as high-purity single photon generation are prerequisite \cite{Chunnilall:14}.
Commonly used single photon counters like avalanche diodes and superconducting nanowires have a short period of breakdown after a detection event \cite{Cova:96,Kerman:SucDetDeadT06}, and require post-corrections to represent real radiant flux with pre-assumptions about photon statistics.
Recent correction techniques supposing a Poisson distribution have achieved repeatable accuracies with parameters of quantum efficiency and detection dead time, promising that the counting-based standard can soon compete \cite{Bae_2019}.
To improve even further, the goal has become achieving single photon generation that deterministically furnishes a single photon only when the detector is ready to count, or at least with a certain and uniform probability \cite{Rodiek:17,Vaigu:2017,Molecule:19}.

Such an ideal source has to maintain single photon flux with low fluctuations and high repeatability.
Current sources, however, show complex behaviors of relaxations, and as a result their emission blinks \cite{Jeong:GAN,Konthasinghe:19}.
In this study, we focus on materials that emit single photon fluorescence at room temperature by investigating silicon vacancy in diamond, defects in GaN, and vacancy in hBN as spectrally narrow and accessible platforms.
We characterize photon number statistics and fluctuations since their degrees are the major factors in determining the accuracy of photon flux for radiometry applications.
Additionally, we compare the maximum count rate allowed for the bare materials under the conventional collection technique of confocal microscopy.
The reason why we stress this condition is that detection count rates are not intrinsic but rather depend on efficiencies given by refractive index geometries and detection techniques.
We note that the present experiments are limited by estimations of internal quantum efficiency and the theoretically maximum count rates under continuous wave operation; more specific methods to optimize collection efficiency, an important subject of photonics, remain for future, application-oriented studies.

To find general tendencies and characteristics from among the complexity and variety of our materials, the current work was based on a large amount of data collected from numerous emitters.
Our full dataset consists of two levels: the first concerns basic properties in identifying single photon emitters, and the second concerns figures of stability.
Data fields in the first level are of photon coincidence correlation $g^{(2)}(0)$ and spectra, which have been used to authenticate single photon fluorescence.
Statistical distributions of the positions of the spectral peaks were collected as a subject for later studies on defect states and their formations.
We discuss the results of these basic properties in \sref{sec:spe}.

Data fields in the second level include photon number uncertainty and the repeatability of measurements for conversion to radiometry flux.
They are examined with a dual detection system that measures photon streams with two modes of detection, namely, photon counting detectors with results in count per second (cps) and photocurrent-generating photodiodes with results in joules per second (W).
Having two different detection mechanisms enables a comparison of outcomes for the same single photon stream as well as an examination of the uncertainty of conversion between the two measures.
Setting the system to photon counting detectors, we measured both photon number fluctuation and repeatability of photon flux measurements to evaluate the degrees of stability both for the photon sources and for the detection system itself, respectively.
Analyzed results are discussed in sections~\ref{sec:stab}--\ref{sec:radiant}.

The main difficulty in radiant flux measurements is to find an emitter that produces a photon flux intense enough to be detected by the photocurrents of the photodiode.
To this end, we exploited an emitter of count rate $> 10^6$ per second and $g^{(2)}(0) < 1$.
We verified both the repeatability of measurement of the radiant flux generated by the single photon emitters and the equivalence of the calibration parameters of detection, as described in \sref{sec:radiant}.
We did not find additional corrections other than our given parameters \cite{Bae_2019}, which was expected from our collection efficiency of $< 6.4 \%$ that flattens any dissimilar photon number statistics into Poisson distribution \cite{Loudon:105699}.
Increasing collection efficiency will be our main subject in future studies to achieve the advantages of sub-Poisson statistics, which is being pursued for few-photon metrology.

\section{Samples and Experimental Methods}\label{sec:method}
Materials of interest are silicon vacancy in diamond (vc-SiV), crystal defects in GaN (df-GaN), and vacancy in hBN (vc-hBN).
Their common feature is a wide band-gap of the host crystal that preserves single photon emission at room temperature.
The SiV sample in this work took the form of nano-diamonds on an iridium substrate, which were grown to the shape of the film by the silicon-free CVD process and milled to diameters of 50--100 nm.
Silicon was implanted after the clean CVD process to attain a high SiV purity \cite{Neu_2011}.
The GaN substrates are a commercially available 4 $\mu$m GaN crystal grown on sapphire \cite{GaN:Spec}.
The fluorescence center in GaN was explained to be a charge-trapped dot that locates at the intersection of a lattice dislocation stemming from the sapphire--GaN interface and a layer mismatch of crystal orientation, similar to the point-like potential wall in a two-dimensional quantum well \cite{Jeong:GAN}.
The hBN sample took the form of nano-flakes dispersed on an oxidized layer of silicon substrate. The nano-flakes are commercially available, but required a special treatment of annealing in an inert environment: 800$^{\circ}$C for 30 min in 1 Torr Ar gas \cite{hBN:ACS2016}.

Because emitters are randomly distributed, confocal microscopy has been commonly used to confine fluorescence signals.
We were helped by the measurement system of our previous work, where we set up modules of single photon collection with single mode fibers (SMFs) and analyzed their photon number statistics \cite{Lim:19}.
The setup has benefits of uninterrupted serial measurements of spatial positions, spectra and photon statistics ($g^{(2)}(0)$), and a high stability of maintained alignments that leads to repeatable measurements (see \sref{sec:radiant}).
The SMF interface gives identical beam profiles for analyzing the modules, leaving the external systems available for exploitation.
The theoretically maximum collection efficiency from the sample--air interface to the SMF output is 21 \%, assuming a Gaussian beam as in past work \cite{Lim:19}.
The collection efficiency $< 6.4$ \% was drawn with consideration of the mode coupling efficiency of an electrical dipole radiation and a SMF mode.\cite{novotny_hecht_2012,Schneider:18} (For more details, see \ref{sec:apenA}.)
The real collection efficiency is smaller than this prediction though when we take into account surface scattering and all the variable nano-crystal shapes.
Still, our photon count rate results are similar to other works that studied the same materials, implying a sufficient level of mechanical rigidity to maintain the count rate.

For the evaluation of photon number fluctuation and application to radiometry experiments, we constructed a radiometry module that includes the dual detection system described above.
This new module has two stages of detection: the first is counting photon flux with a silicon single photon avalanche detector (SPAD), and the second is measuring radiant power converted from the photodiode photocurrents.
They share the same single photon input injected via SMF, and the same incidence position at which we rotate groups of detectors to place.
This intends to attain convergence between the outcomes of the two detection mechanisms consistently without any corrections for optical path loss (\sref{sec:radiant}).
The module also helped the measurement of photon number fluctuations, as shown in \sref{sec:stab}.


\section{Spectra and Photon Statistics}\label{sec:spe}

\begin{figure}[ht!]
\begin{center}
\includegraphics[width=10cm]{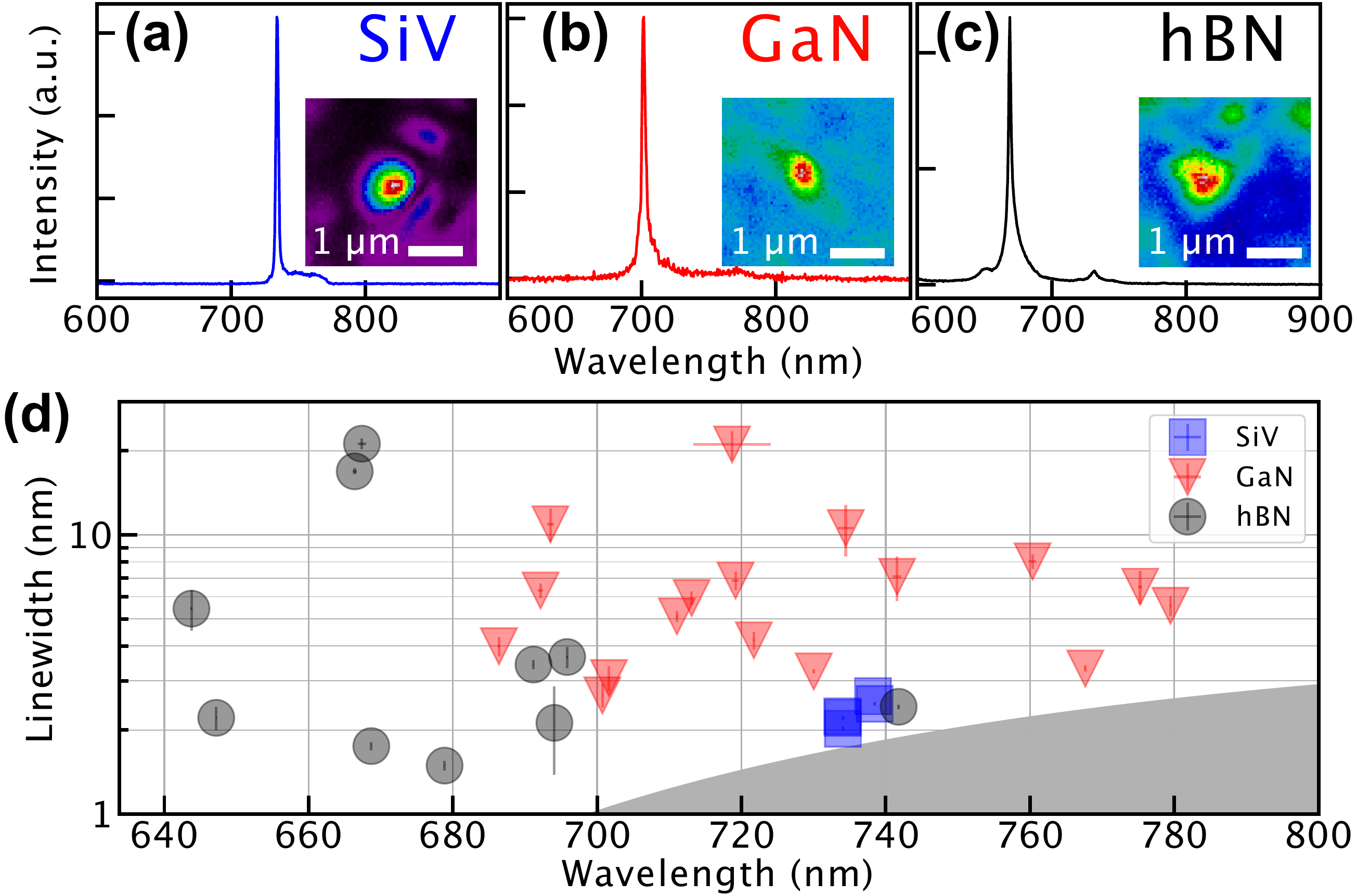}
\end{center}
	\caption{Spectra of single photon emitters of different materials: \textbf{(a)} silicon vacancy in diamond (SiV), \textbf{(b)} fluorescence defect in GaN crystal (GaN), and \textbf{(c)} vacancy in hBN (hBN). Insets are photoluminescence images scanned over the area centered on each fluorescence center.
	\textbf{(d)} Histograms of spectral peaks of wavelength and linewidth for SiV (blue), GaN (red), and hBN (gray).
	The gray zone is the resolution boundary limited by our spectrometer and its coupling optics.
	}
	\label{fig:spectra}
\end{figure}

Spectra of fluorescence centers commonly consist of a zero-phonon line (ZPL) and phonon side-bands, and show a high ratio of ZPL intensity compared to the photon side-band, i.e., a high Debye--Waller factor, as shown in \fref{fig:spectra}(a--c).  
The ZPL positions of vc-hBN and df-GaN depend on strain and defect formations, and are widely distributed over 600--750 nm and 690--800 nm, respectively [\fref{fig:spectra}(d)].
Due to the different mechanisms of defect formation between df-GaN and vc-hBN, df-GaN has a greater linewidth on average and can be more directly affected by crystal strain.
Likewise, both df-GaN and vc-hBN have a wider linewidth on average, which can vary by the local strain of the host crystal because of the various kinds of defect formations and their large degrees of freedom.
On the other hand, vc-SiV has a definite ZPL position, $\sim737$ nm\cite{SiV:1996}, the formation of which is explicitly allowed by the diamond crystal.

\begin{figure}[t]
\begin{center}
\includegraphics[width=12cm]{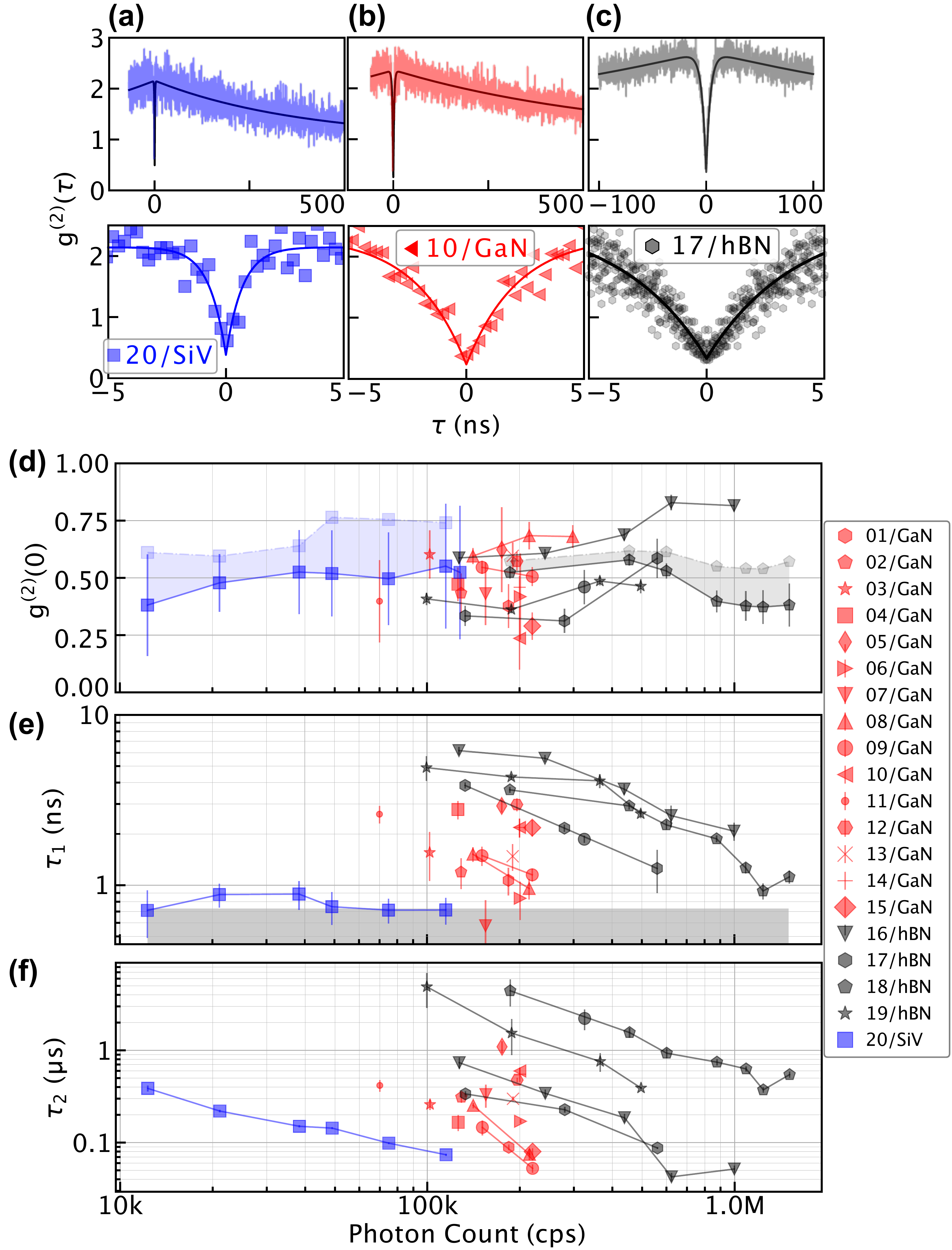}
\end{center}
\caption{Photon correlation ($g^{(2)}(\tau)$) acquired from a Hanbury Brown--Twiss interferometer for \textbf{(a)} silicon vacancy in diamond (blue), \textbf{(b)} defects in GaN (red), and \textbf{(c)} vacancy in hBN (gray). Solid lines show the model $g^{(2)}(\tau) = 1 - p_1 \exp (-\tau/\tau_1) + p_2 \exp (-\tau/\tau_2)$ with characteristic times of anti-bunching ($\tau_1$) and trapping in meta-stable dark states ($\tau_2$).
Fitted to this model, the zero-time correlations ($g^{(2)}(0)$) are derived to be \textbf{(a)} 0.38 $\pm$ 0.22, \textbf{(b)} 0.24 $\pm$ 0.14,  and \textbf{(c)} 0.33 $\pm$ 0.05 (95 \% trust).
From the full set of data attained from various fluorescence centers in the three materials, photon counts as acquired from the detectors are plotted for \textbf{(d)} $g^{(2)}(0)$, \textbf{(e)} $\tau_1$, and \textbf{(f)} $\tau_2$.
The gray zone in the figure (e) is the resolution boundary limited by time jitter noises of single photon detectors.
}
\label{fig:g2}
\end{figure}

For a statistical approach, we collected photon statistics from $>$ 20 fluorescence emitters.
Single photon emitters have a unique property of fluorescence, exhibiting a low coincidence of photon count.
The degree and time scale of this coincidence suppression have been commonly represented by a normalized correlation ($g^{(2)}(\tau) = \langle C(t+\tau) C(t)\rangle/\langle C(t)\rangle^2$) of photon count ($C$) as measured by a Hanbury Brown--Twiss interferometer (HBT) \cite{Loudon:105699}.
We employed two methods of deducing $g^{(2)}(\tau)$ experimentally: start-stop histogram and time-tag correlation (TTC).
The former is advantageous for real-time acquisition as the trigger intervals between signals from two SPADs are collected.
The latter though has a better convergence at $g^{(2)}(\infty) \rightarrow 1$ because this method stores time-tags crudely and deduces a normalization factor from the total count of each detector during data processing, 
which gives reliable results of $g^{(2)}(0)$.
In our study, estimations of $g^{(2)}(\tau)$ were based on raw data given by the TTC method.

The $g^{(2)}(\tau)$ from our samples, however, shows composite features of both anti-bunching ($g^{(2)}(\tau) < 1 $) at $|\tau| < \tau_1$ and bunching ($g^{(2)}(\tau) > 1$) at $\tau_1<|\tau| < \tau_2$ simultaneously.
Here, $\tau_1$ and $\tau_2$ are effective time constants defined by the fitting model $g^{(2)}(\tau) = 1 - p_1 e^{-|\tau|/\tau_1} + p_2 e^{-|\tau|/\tau_2}$, where $p_1$ is the depth contributing to the anti-bunching of $g^{(2)}(0)$, $p_2$ is the height above $g^{(2)}(\tau) > 1$,
$\tau_1$ is the time width of anti-bunching, and $\tau_2$ is the characteristic exponential decay time of bunching.
We collected $\tau_1$ and $\tau_2$ because they have physical origins: $\tau_1$ is the spontaneous emission time for an ideal photon source or relates to the lifetime of the radiative transition for single photon emission, while $\tau_2$ stems from non-radiative relaxations, which can cause blinking in the photon emission.\cite{Santori:2004bv}
If a single emitter is trapped in meta-stable dark states, it stops emitting until it is released to bright states, with $\tau_2$ directly representing this time scale.

With this model, we observed $g^{(2)}(0) = 0.38 \pm 0.22$ for vc-SiV, $0.24\pm 0.14$ for df-GaN, and $0.33 \pm 0.05$ for vc-hBN, as shown in \fref{fig:g2}(a--c), where the errors are the widths of the 95 \% confidence interval calculated via robust covariance estimation \cite{RCVME}. 
The full $g^{(2)}(0)$ data was measured from $>$ 20 fluorescence centers of our samples as shown in \fref{fig:g2}(d).
The large errors in the $g^{(2)}(0)$ obtained from SiV emitters are due to the short $\tau_1$ of these emitters and a time jitter of the detector.
To restore the pure $g^{(2)}(0)$ before time jitter noises, we tried a deconvolution method with a noise filter  $H(\tau;D) = (D\sqrt{2 \pi})^{-1}\exp(-\tau^2/2 D^2)$ assumed for a time jitter $D = 0.3$ ns.\cite{Nomura:2010gx}
Our method of deconvolution is to fit data with a convolution form of exponential function.
However, this method is redundant as its results are similar to the previous estimation of $g^{(2)}(0)$ with parameters of $p_1$ and $p_2$, and they can be biased when $D$ is overestimated.
It is a common tendency regardless of materials that time jitter errors of $g^{(2)}(0)$ grows as $\tau_1$ is shortened.
Since $g^{(2)}(\tau)$ measured by TTC is in absolute values, we could take the average $g^{(2)}(0)$ for some intervals of $\tau \in (-\tau_1, \tau_1)$.
These experimentally allowed values are also presented as transparent points from 18/hBN fluorescence emitter shown in \fref{fig:g2}(d).
Their differences from pure values $g^{(2)}(0)$ of the model are reflected in confidence intervals for all emitters.

Our closest recorded $g^{(2)}(0)$ values to 0 were attained around $0.24\pm 0.14$ from df-GaN and $0.31 \pm 0.05$ from vc-hBN, both of which are allowed at a low excitation power.
The effective excitation power differed among materials and samples due to the differences in refractive index, diameter of grains, and geometry.
However, the set of data used for \fref{fig:g2}(d--f) comprises the full range of photon count rates achieved from far below and above the saturation points of each material.
We suspect that the lowest value of $g^{(2)}(0) > 0.2$ can either be attributed to background photoluminescence, including deeper infrared, as we used a long pass filter with a 568 nm edge for purification, or the unresolved single photon emitting transitions in a ZPL \cite{Alexander:2019um}.
We discuss this in \sref{sec:radiant} with \fref{fig:conv}.

Our result shows that $\tau_1$ and $\tau_2$ decrease with an increasing excitation power for every material we observed.
These variables of the $g^{(2)}(\tau)$ model has an origin in relaxation processes of fluorescence materials.
The power dependence of $\tau_1$ evidently shown with df-GaN and vc-hBN implies that $\tau_1^{-1}$ represents a re-excitation rate rather than a spontaneous emission rate under moderate $P$ that allows exploitable photon count rates.
Nevertheless, we can expect large spontaneous emission rates from SiV, whose $\tau_1$ is as short as being close to the instrumental limit of time jitter at the entire range of $P$.
According to the three level model, $\tau_2$ are related to a recovering relaxation from meta-stable states (deshelving), and also depends on excitation power because it gives more chances of initializing ionization \cite{Santori:2004bv,Jeong:GAN,ASTAKHOV2018211}.

We observed high count rates of $> 10^6$ cps with vc-hBN, similar to other studies \cite{C7NR08249E}, and low count rates of $< 2\times 10^5$ cps with vc-SiV.
This result is opposed to the long $\tau_{1,2}$ of hBN, as shown in \fref{fig:g2}(e) and (f), and to the intuition that fast transitions allow high photon rates.
The speed of the transitions and blinking of vc-SiV was the highest among the materials of interest, but exact pictures of these have yet to be unveiled to predict the internal efficiency of fluorescence emissions at room temperature \cite{SiV:APL:2011,Lindner_2018}.
The photon count rate of df-GaN, $< 3 \times 10^5$ cps, seems limited by total internal reflections at the GaN--air interface, which can be overcome by an immersion medium before the objective lens.
Otherwise, vc-hBN is preferable for radiometry experiments that require a wide range of photon counts on the order of $> 10^6$ cps.


\section{Photon number fluctuation}\label{sec:stab}

\begin{figure}[h]
	\centerline{\includegraphics[width=7.5 cm]{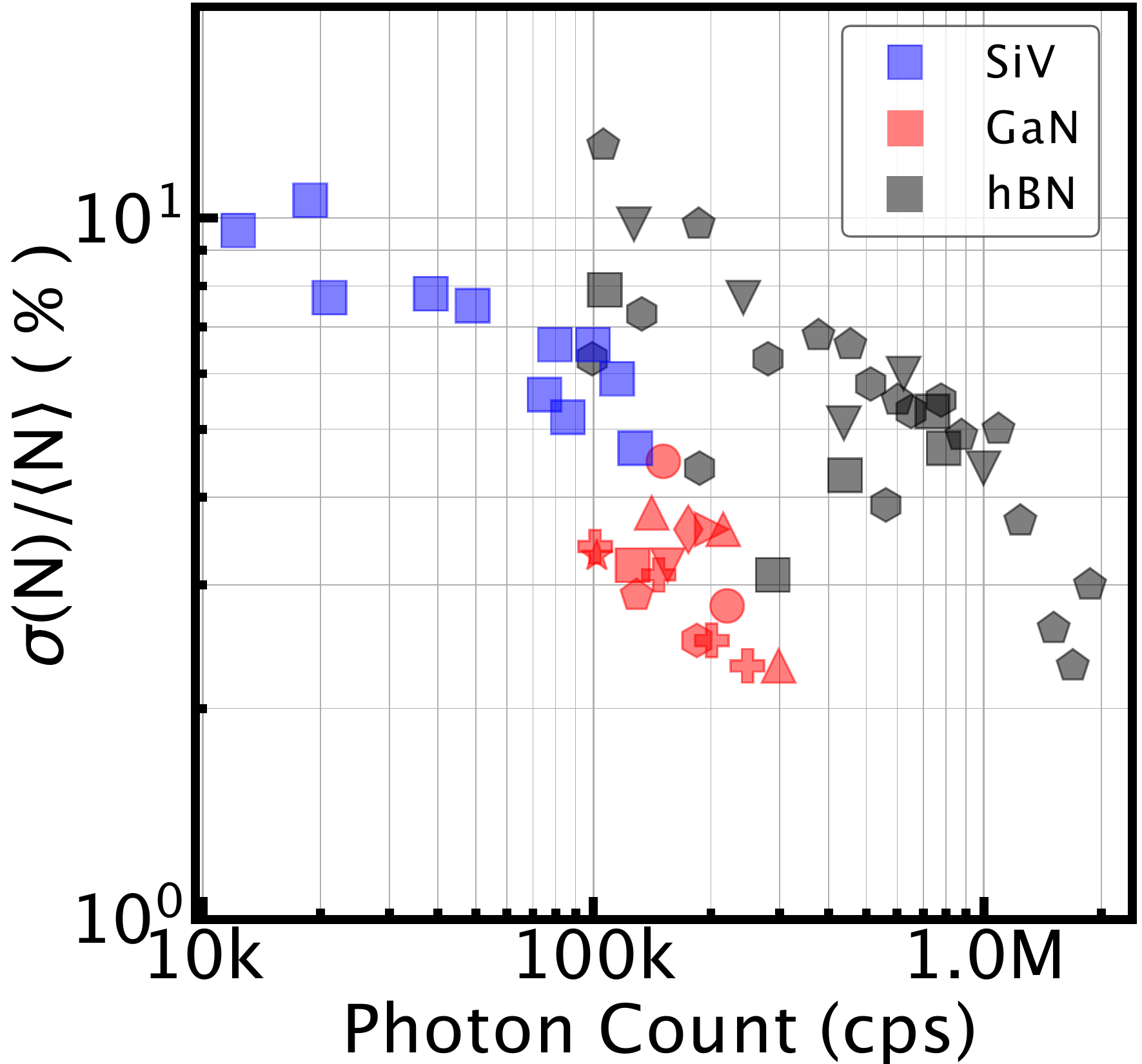}}
	\caption{Relative number uncertainty $\sigma(N)/\langle N\rangle$ of average photon number $\langle N\rangle$ from various fluorescence centers of different materials: silicon vacancy in diamond (SiV, blue), defect in GaN (red), and vacancy in hBN (gray).
	Each point was acquired from a 10 s long streaming acquisition with a time bin width of 10 ms.
	Data collected with different emitters were distinguished by shape of marker specified in \fref{fig:g2}.
	}
	\label{fig:stability}
\end{figure}

Low photon number uncertainty is required for an ideal photon source.
From various values of $\tau_2$ depending on the emitters and materials, which are related to the time scales of emission blinking, we can infer the requirement to find one that has a low photon number fluctuation.
We first measured the photon number uncertainty ($\sigma (N)/\langle N\rangle$), defined as the ratio of the standard deviation of photon number ($\sigma (N)$) to its average value  ($\langle N\rangle$).
These statistical variables were obtained by a single shot of streaming acquisition of photon counts over 10 s.
Photon number $N$ is the accumulated photon count in a 10 ms long time bin ($\Delta t$) within the streaming acquisition.
Thus, we clarify the terminological distinction between $N$ and photon count rate $C$ by the following relation:
\begin{equation}
N_i = \int_{t_i}^{+ \Delta t}  C(t)\,  \mathrm{d}t,
\end{equation}
for an $i$-th time bin $[t_i, t_i + \Delta t]$.
In experiments, we measured $N_i$ directly from an edge counter with a fixed $\Delta t$ and then deduced $C_i$ via $C_i = N_i/\Delta t$.

For shot-noise limited $N$ in Poisson statistics as acquired for $\Delta t > \tau_1$, the defined photon number uncertainty $\sigma (N)/\langle N\rangle$ can be reduced to $1/\sqrt{N}$, which tends to decrease with increasing $N$.
This assumption excludes the observed differences of uncertainty between the materials shown in \fref{fig:stability}(a).
Measured values of the uncertainty are split into two groups: SiV with GaN, and hBN on the other side.
With a few exceptions, every hBN emitter exhibited a larger $\sigma (N)/\langle N\rangle$ than SiV and GaN.

\begin{figure}[h]
	\centerline{\includegraphics[width=10cm]{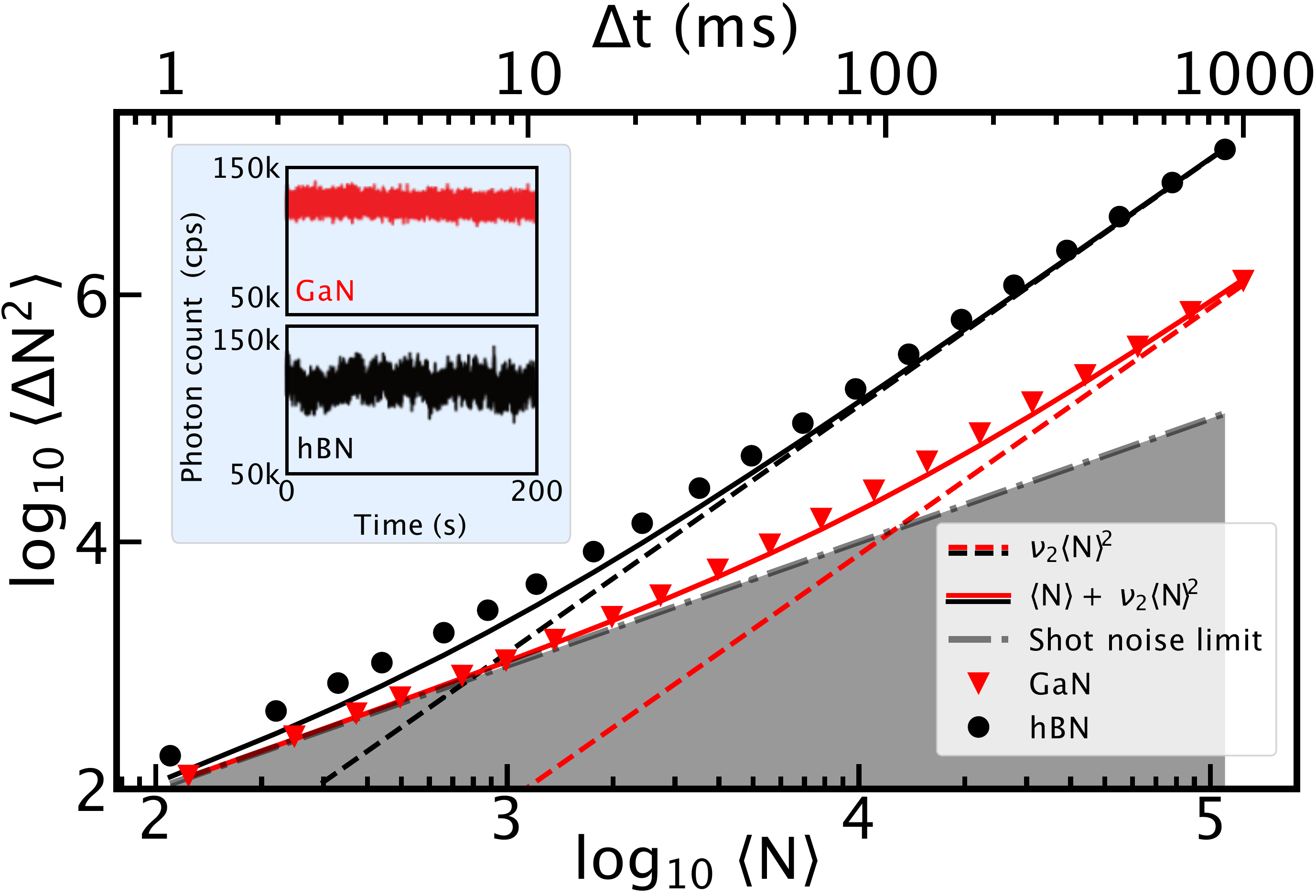}}
	\caption{Photon number variance $\langle \Delta N^2\rangle$ with respect to average photon number $\langle N \rangle$, measured from fluorescence emissions of defect in GaN (red) and vacancy in hBN (black) samples.
	Photon number $N$ is an integrated variable of count rate $C$ for time bin $\Delta t$ as $N = C \Delta t$.
	Varying $\Delta t$ from 1 ms to 1 s, the statistical values of $N$ were taken from streaming acquisitions for 200 s, as shown in the inset.
	The gray line is the shot-noise limit: $\langle \Delta N^2\rangle_{\mathrm{shot}} = \langle N \rangle$.
	The solid lines follow the model $\langle \Delta N^2\rangle = \langle N \rangle + \nu_2 \langle N \rangle^2$, with $\nu_2 = 7.9 (\pm 0.3)\times10^{-5}$ for GaN and $1.3 (\pm 0.05)\times10^{-3}$ for hBN.
	}
	\label{fig:fluc}
\end{figure}

Additional noises other than the shot-noise $\langle \Delta N^2\rangle_{\mathrm{shot}} = \langle N \rangle$ are clearly seen.
We examined $\langle \Delta N^2\rangle$ with integration times ($\Delta t$) from 1 ms to 1 s for df-GaN and vc-hBN;
$\langle N \rangle$ was set to a similar value ($1.3 \times 10^5$ cps) by adjusting excitation power.
In this condition, $g^{(2)}(0) = 0.41 \pm 0.04$ for the df-GaN and $0.54 \pm 0.03$ for the vc-hBN, with both emitters below saturation.
We extrapolated the noise model to include the quadratic term of $\langle N \rangle$ with the coefficient $\nu_2$:
\begin{equation}
\langle \Delta N^2 \rangle = \langle N \rangle +  \nu_2 \langle N \rangle^2.
\end{equation}
This model was fitted to df-GaN with $\nu_2 = 7.9 (\pm 0.3) \times 10^{-5}$, and as shown in \fref{fig:fluc},  $\langle \Delta N^2 \rangle$ was close to $\langle \Delta N^2\rangle_{\mathrm{shot}}$ at $\Delta t <$ 20 ms and small $\langle N \rangle$.
However, the data from vc-hBN was close to the model only for large $\langle N \rangle > 10^4$ with $\nu_2 = 1.3 (\pm 0.05) \times 10^{-3}$, and did \textit{not} converge to $\langle \Delta N^2\rangle_{\mathrm{shot}}$ at short $\Delta t$.
The $\langle \Delta N^2 \rangle$ of vc-hBN was greater than that of df-GaN by an order of magnitude, as much as the difference of $\nu_2$ between them.

The large fluctuation of $N$ in vc-hBN can easily be seen from the real-time data in the inset of \fref{fig:fluc}.
The origin is related with the blinking phenomenon, as vc-hBN has long $\tau_2 = 1.0 \pm 0.1$ $\mu$s.
This value is in contrast to the short $\tau_2= 53 \pm 5$ ns of df-GaN.
In our survey of $>$ 36 fluorescence centers, most vc-hBN $\tau_2$, which are related to the time dynamics of the blinking, are slower than those of GaN, as shown in \fref{fig:g2}(f).
This analysis of $\tau_2$ is still insufficient to extrapolate the general blinking dynamics though, since we could not measure $\tau_2 > 1$ ms $\sim \Delta t$ in $g^{(2)}(\tau)$ due to the sampling time window.
Nevertheless, this tendency of slow blinking dynamics being more clearly seen in vc-hBN, as shown in the inset of \fref{fig:fluc}, is intriguing.
On the other hand, df-GaN with the fast $\tau_2$ did \textit{not} exhibit such fluctuations, and enabled the low noise $\sim \langle \Delta N^2\rangle_{\mathrm{shot}}$ with $\Delta t < 10$ ms.

Because of the low $\langle N \rangle$ of vc-SiV, we could not perform a fair test of $\nu_2$ for the vc-SiV samples.
However, the analogous behavior of $\langle \Delta N^2 \rangle$ with df-GaN as shown in \fref{fig:fluc} leads us to expect that $\langle \Delta N^2 \rangle$ of vc-SiV would be as low as that of df-GaN if it had a similar level of $\langle N \rangle$.
The low $\langle N \rangle$ of vc-SiV also has disadvantages in obtaining repeatable $S$, as in the following section.


\section{Conversion Between Photon and Radiant Fluxes with High Repeatability}\label{sec:radiant}

In order to apply single photon emitters in radiometry experiments, evaluations of reliability in measurement results must precede.
The main quality of our assessment is the repeatability of photon flux ($\Phi_q$) or photon count ($C$) measurement followed by conversion to radiant flux ($S$).
We performed two tests: one to deduce the repeatability errors of $\langle N\rangle$, and the other to confirm the validity of applying present calibration parameters for photon counts to represent radiant fluxes.
The dual detection module introduced in \sref{sec:method} is well suited for performing these tests.
With this module, we measured $C$ and $S$ from a SPAD and photodiode, respectively, and cross-checked the independent results with previously proven calibration parameters.

\begin{figure}[h]
	\centerline{\includegraphics[width=7.5 cm]{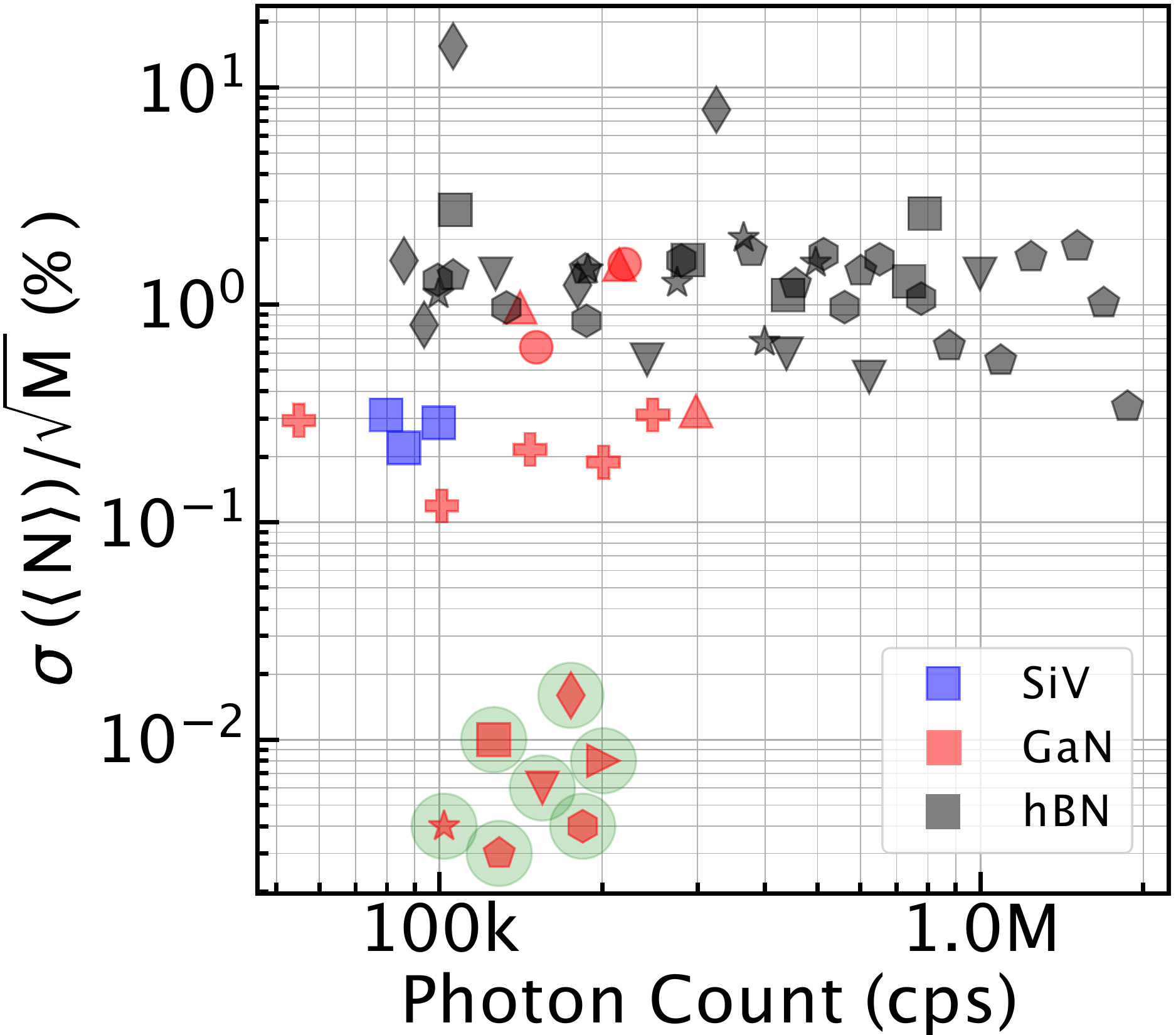}}
	\caption{Repeatability of $\langle N\rangle$ ($=\sigma(\langle N\rangle)/\sqrt{M}$, where $M$ is the number of repeated measurements). 
	Single values of $\langle N \rangle$ were acquired from a 10 s long streaming acquisition at a sampling rate of 10 ms.
	$\sigma(\langle N\rangle)$ was calculated by $M$ values of $\langle N \rangle$, where $M=5$ was set to extract a quantity of repeatability.
	Data in the green circles were attained from high repetitions up to $M=20$.
	Data collected with different emitters were distinguished by shape of marker specified in \fref{fig:g2}.
	}
	\label{fig:repeat}
\end{figure}

Repeatability errors were measured via the following process.
We repeated the streaming acquisition of $N$ the same as we did for photon number uncertainty in \sref{sec:stab}.
We took the average value ($\langle N\rangle$) of each shot ($i$) and repeated $M$ times to obtain $\langle N\rangle_{1\leq i\leq M}$, thereby yielding the repeatability error as $\sigma (\langle N \rangle)/\sqrt M$ according to its theoretical definition.
In order to obtain practical values, we inserted an $S$ measurement process between each shot of $N$ acquisition to mirror the calibration sequence in radiometry experiments.
Hence, we attain results of both repeatability error and data on $C$ and $S$.

As shown in \fref{fig:repeat}, df-GaN, which has low $\langle \Delta N^2 \rangle$, demonstrates a high repeatability of $\langle N\rangle$ measurement.
We extended $M$ to 20 to obtain the upper bound of repeatability with a qualified df-GaN emitter with $g^{(2)}(0) = 0.43 \pm 0.09$ and $C = 1.3 \times 10^5$ cps.
This highest result reaches to 30 ppm, as marked with green circles in \fref{fig:repeat}.
We note that this value is close to the present repeatability of radiometry experiments with this laser source.

\begin{figure}[h]
	\centerline{\includegraphics[width=10cm]{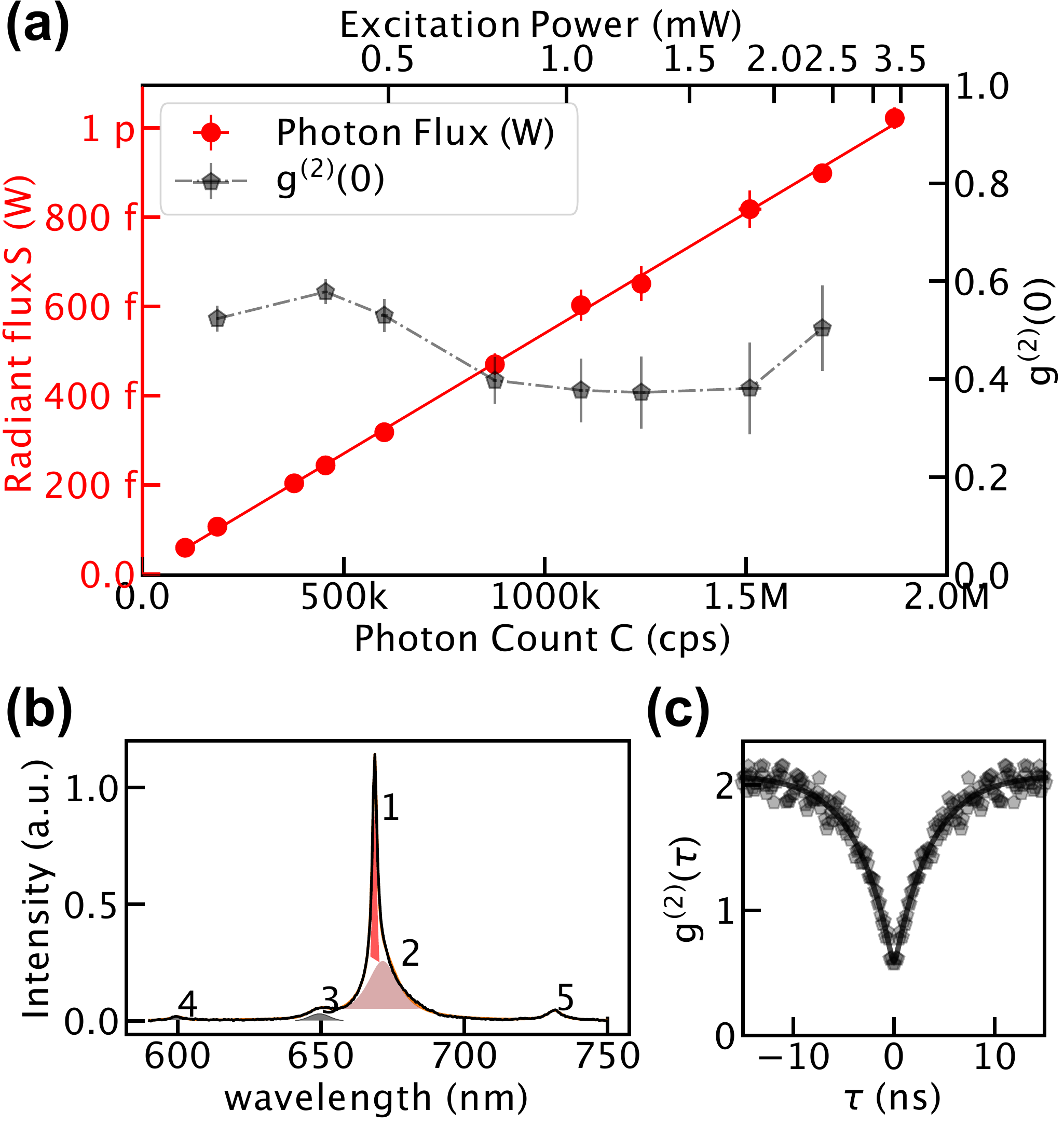}}
	\caption{\textbf{(a)} Relation of radiant flux (red dots, $I$) with respect to photon count rate ($C$), where detections are based on different mechanisms of operation.
	The radiant fluxes (y-axis) are given by currents produced in a traceable, high-sensitivity photodiode, and the photon fluxes are from a single photon counter (SPC).
	The solid red line represents the theoretical relation $S = \frac{hc}{\lambda} \eta \Phi_q$ (identical to \eref{eq:identity}), where $\eta$ is the quantum efficiency of the SPC and $\lambda$ is the center wavelength.
	The black squares show $g^{(2)}(0)$ varied at different levels of $C$.
	\textbf{(b)} The spectrum of the single photon source presents mainly as the Lorentzian function centered on 665 nm with a FWHM of 2 nm in wavelength (1) while containing small peaks (2--5).
	 \textbf{(c)} Photon autocorrelation measured at an excitation power of 0.2 mW and a photon count of $2 \times 10^5$ cps.
	 All were measured with 18/hBN (see \fref{fig:g2}).
	}
	\label{fig:conv}
\end{figure}

Despite a low fluctuation and high repeatability, the maximum $C$ obtained from df-GaN has a lower limit around $2 \times 10^5$ cps [\fref{fig:g2}(d--f)].
This also limits the signal to noise ratio (SNR) to 22 for $S$ measurements by a silicon photodiode having a noise equivalent power (NEP) of 8.4 fW/$\mathrm{Hz}^{\frac{1}{2}}$ and integrating photocurrents for 10 s \cite{Mountford:08,Park:16}.
To take advantage of the high SNR of $S$ and the wide range of flux levels shown in \fref{fig:conv}, we chose 18/hBN among vc-hBN emitters (see \fref{fig:g2}), whose maximum $C \sim 2 \times 10^6$ cps.
We adjusted the level of $C$ by controlling the excitation power ($P$).
$C$ has a behavior following the function $C =C_0 (1 - e^{-P/P_\mathrm{sat}})$ with a saturation count rate of $C_0 = 1.8\times 10^6$ cps and a saturation power of $P_\mathrm{sat} = 1.3 \pm 0.2$ mW.
Over this wide range of $C$, $g^{(2)}(0)$ remained within the range 0.37--0.58.
The uncertainty of $g^{(2)}(0)$ (confidence interval 95 \% trust) increased at high $C$, as strong excitation power shortens the anti-bunching time width $\tau_1$ into the time jitter limit of SPAD, as shown in \fref{fig:g2}(e).

As shown in \fref{fig:conv}(c), $g^{(2)}(0) = 0.53 \pm 0.03$ when measured at a low excitation power of $0.15 \times P_\mathrm{sat}$, and it decreases to 0.35 at higher $P$.
We attribute this irregularity to the presence of other transitions that independently emit single photons.
Spectral investigations reveal that the ZPL is composed of two Lorentzian peaks, labeled \textit{1} and \textit{2} in \fref{fig:conv}(b).
Similar findings were observed in another work, and according to the previous analysis we can predict that cross-correlations between independently emitted single photons increase $g^{(2)}(0)$ \cite{Alexander:2019um}.
As the overlapped peaks \textit{1} and \textit{2} have mostly similar areas, the degree of mixture is sufficient to be the main cause of the high $g^{(2)}(0)$.
We also attribute the oscillating behavior of $g^{(2)}(0)$ with respect to $P$ to a property of mixture in which different excitation cross-sections of fluorescence transitions make them compete in contributing to a sum of photon count rate.

Thanks to the narrow linewidth ($\Delta \lambda$ $\sim 2$ nm) shown in \fref{fig:conv}(b), we took a single value of $\eta$ at the center wavelength $\lambda_c = 665.8 \pm 0.03$ nm \cite{Bae_2019}, with the justification that $\Delta \eta$ at the $\lambda$ within the narrow line is smaller than the measured uncertainty (0.5 \%).
Error caused by detection dead time and nonlinear counting was predicted to be 0.2 \% at $C=10^5$ cps and grow to 4 \% at $C=5 \times 10^5$ cps, so that error correction is critical for $C > 10^5$ cps.
For this, we used the correction function $\hat C = U(C)$ to present the corrected count rate $\hat C$, as defined in a previous work \cite{Bae_2019}.
At this stage, with given quantum efficiency $\eta$, we can extract the important $\Phi_q = \eta \times \hat C$.
In practice, we can reduce this to $\Phi_q = \mathrm{DE} (C) \times C$ by introducing an effective function $DE (C)$ that includes both effects of $\eta$ and $U: C \rightarrow \hat C$ \cite{Bae_2019}.
Then we represent the relation between $S$ and $\Phi_q$ in a convenient form with the uncorrected variable $C$:
\begin{equation}\label{eq:identity}
S = \frac{hc}{\lambda} \times \mathrm{DE} (C) \times C.
\end{equation}
This relation with given $\mathrm{DE} (C)$ agrees well with the experimental $S$ and $C$ data shown in \fref{fig:conv}(a).
We first note that the parameters given in previous works were applied here identically, even though the measurement system in this work was newly fashioned.
Second, identical results were achieved even with a non-classical light source that was not fully controlled, as this new source contained uncontrolled internal dynamics related to $\tau_2$ and the blinking phenomenon.
Such a photon source may reduce the repeatability of $C$ due to a high $\langle\Delta N^2\rangle$, but it does not cause contradictions with the given parameters in $\mathrm{DE} (C)$ because the parameters are central values.




\section{Discussions}

We have studied single photon emitters based on fluorescence defects in various crystal materials at room temperature.
Silicon vacancy in diamond, defects in GaN, and vacancy in hBN platforms all have a narrow linewidth, and their spectrum centers are spread over a wide range of wavelengths.
This is, in fact, an important advantage for few-photon radiometry where we have various quantum efficiencies as variable parameters in conversion models for radiant power.  

Both stability and brightness are important qualities of single photon sources in few-photon metrology.
None of the materials investigated in this study are endowed with both characteristics simultaneously.
For example, although single photon sources in hBN nano-flakes exhibit high rates of photon detection, they are interrupted by slow blinking, which causes severe fluctuations in photon count rates.
On the other hand, those in GaN show a high degree of stability and a high repeatability of emission rate, while their photon count rates are lower than those attained from hBN.

Such differences can be expected from their shapes.
Emitters in hBN are close to the surface or edges in a nano-flake, where electrostatic fields can have large effects and cause vulnerability to charge fluctuations.
The scenario of severe blinking in hBN is also supported by a recent study that revealed many internal states and frequent relaxations between them, even at low temperature.
On the other hand, emitters in GaN are embedded at a depth of a few micrometers, and this reduces the fluctuations caused by electrostatic fields.
However, the flat surface with high refractive index of GaN significantly decreases its photon collection efficiency, according to the total internal reflection effect \cite{Bowman:14}.

Various methods have been developed to alleviate total internal reflection.
This has long been a subject to mitigate the surface effects in solar cells and light-emitting diodes, and many techniques have been developed.
The simplest  solution is a solid immersion lens that includes various lens techniques like micro-half spheres and meta-surfaces.
These methods do not modify the regions near the emitters beneath the surface.
We are expecting high collection efficiencies as supported by these methods, which will enable us to apply non-classical photon number statistics, like one corresponding to the Fock state, and to achieve a high degree of number uncertainty.

\appendix
\section{Coupling efficiency of radiation from a point dipole source to a single mode optical fiber}\label{sec:apenA}
We assume an electric dipole source laid at a focal point in air and perpendicular to cylindrical symmetry axis connecting an objective lens and a single mode fiber (SMF) collimation lens.
Electromagnetic fields that refocused by the SMF collimation lens were analytically calculated from the Green dyadic function,
while the radio of field intensity collected through a finite numerical aperture 0.95 ($\eta_\mathrm{NA}$) was extracted \cite{novotny_hecht_2012}.
Because the refocused field has a different profile from a SMF mode, a field-to-field coupling efficiency ($\eta_\mathrm{SMF}$) is $< 73.7 \%$ calculated using an inner product defined in \cite{Schneider:18}.
Then, $\eta_\mathrm{NA} \times \eta_\mathrm{SMF} = 32 \%$.
Considering a setup loss $20 \%$, the overall efficiency excluding unknown substrate losses is reduced to $6.4 \%$.

\section*{Acknowledgments}
This work was partially funded by National Research Council of Science and Technology (NST) of the Republic of Korea (CAP-15-08-KRISS) and Korea Research Institute of Standards and Science (KRISS) (KRISS-2019-GP2019-0020). The authors declare no conflicts of interest.

\section*{References}

\begin{thebibliography}{10}
\expandafter\ifx\csname url\endcsname\relax
  \def\url#1{{\tt #1}}\fi
\expandafter\ifx\csname urlprefix\endcsname\relax\def\urlprefix{URL }\fi
\providecommand{\eprint}[2][]{\url{#2}}

\bibitem{Tittel:01}
Tittel W and Weihs G 2001 {\em Quantum Info. Comput.\/} {\bf 1} 3–56

\bibitem{chen:06}
Chen S, Chen Y~A, Strassel T, Yuan Z~S, Zhao B, Schmiedmayer J and Pan J~W 2006
  {\em Phys. Rev. Lett.\/} {\bf 97}(17) 173004

\bibitem{Willson:73}
Willson R~C 1973 {\em Appl. Opt.\/} {\bf 12} 810

\bibitem{Martin_1985}
Martin J~E, Fox N~P and Key P~J 1985 {\em Metrologia\/} {\bf 21} 147

\bibitem{QCandelta:07}
Cheung J~Y, Chunnilall C~J, Woolliams E~R, Fox N~P, Mountford J~R, Wang J and
  Thomas P~J 2007 {\em J. Mod. Opt\/} {\bf 54} 373

\bibitem{Zwinkels:10}
Zwinkels J~C, Ikonen E, Fox N~P, Ulm G and Rastello M~L 2010 {\em Metrologia\/}
  {\bf 47} R15

\bibitem{Chunnilall:14}
Chunnilall C~J, Degiovanni I~P, Kück S, Müller I and Sinclair A~G 2014 {\em
  Optical Engineering\/} {\bf 53} 1

\bibitem{Cova:96}
Cova S, Ghioni M, Lacaita A, Samori C and Zappa F 1996 {\em Appl. Opt.\/} {\bf
  35} 1956

\bibitem{Kerman:SucDetDeadT06}
Kerman A~J, Dauler E~A, Keicher W~E, Yang J~K~W, Berggren K~K, Gol’tsman G
  and Voronov B 2006 {\em Appl. Phys. Lett.\/} {\bf 88} 111116

\bibitem{Bae_2019}
Bae I~H, Park S, Hong K~S, Park H~S, Lee H~J, Moon H~S, Borbely J~S and Lee D~H
  2019 {\em Metrologia\/} {\bf 56} 035003

\bibitem{Rodiek:17}
Rodiek B, Lopez M, Hofer H, Porrovecchio G, Smid M, Chu X~L, Gotzinger S,
  Sandoghdar V, Lindner S, Becher C and Kuck S 2017 {\em Optica\/} {\bf 4} 71

\bibitem{Vaigu:2017}
Vaigu A, Porrovecchio G, Chu X~L, Lindner S, Smid M, Manninen A, Becher C,
  Sandoghdar V, Götzinger S and Ikonen E 2017 {\em Metrologia\/} {\bf 54} 218

\bibitem{Molecule:19}
Lombardi P, Trapuzzano M, Colautti M, Margheri G, Degiovanni I~P, López M,
  Kück S and Toninelli C 2020 {\em Adv. Quantum Technol.\/} {\bf 3} 1900083

\bibitem{Jeong:GAN}
Berhane A~M, Jeong K~Y, Bodrog Z, Fiedler S, Schr{\"o}der T, Trivi{\~n}o N~V,
  Palacios T, Gali A, Toth M, Englund D and Aharonovich I 2017 {\em Adv.
  Mater.\/} {\bf 29} 1605092

\bibitem{Konthasinghe:19}
Konthasinghe K, Chakraborty C, Mathur N, Qiu L, Mukherjee A, Fuchs G~D and
  Vamivakas A~N 2019 {\em Optica\/} {\bf 6} 542

\bibitem{Loudon:105699}
Loudon R 1973 {\em {The Quantum Theory of Light}\/} (Oxford: Clarendon Press)

\bibitem{Neu_2011}
Neu E, Steinmetz D, Riedrich-Möller J, Gsell S, Fischer M, Schreck M and
  Becher C 2011 {\em New J. Phys\/} {\bf 13} 025012

\bibitem{GaN:Spec}
The film (2-$\mu$m p-type/2-$\mu$m undoped) grown on sapphire, supplied by
  Suzhou Nanowin Science and Technology Co., Ltd..

\bibitem{hBN:ACS2016}
Tran T~T, Elbadawi C, Totonjian D, Lobo C~J, Grosso G, Moon H, Englund D~R,
  Ford M~J, Aharonovich I and Toth M 2016 {\em ACS Nano\/} {\bf 10} 7331

\bibitem{Lim:19}
Lim H~J, Jeong K~Y, Lee D~H and Hong K~S 2019 {\em Opt. Mater. Express\/} {\bf
  9} 4644

\bibitem{novotny_hecht_2012}
Novotny L and Hecht B 2012 {\em Principles of Nano-Optics\/} (Cambridge
  University Press) chap~3 2nd ed

\bibitem{Schneider:18}
Schneider P~I, Srocka N, Rodt S, Zschiedrich L, Reitzenstein S and Burger S
  2018 {\em Opt. Express\/} {\bf 26} 8479

\bibitem{SiV:1996}
Goss J~P, Jones R, Breuer S~J, Briddon P~R and \"Oberg S 1996 {\em Phys. Rev.
  Lett.\/} {\bf 77}(14) 3041

\bibitem{Santori:2004bv}
Santori C, Fattal D, Vu{\v c}kovi{\'c} J, Solomon G~S, Waks E and Yamamoto Y
  2004 {\em Phys. Rev. B\/} {\bf 69} 205324

\bibitem{RCVME}
Sidik K and Jonkman J~N 2016 {\em Statistics in Medicine\/} {\bf 35} 4856

\bibitem{Nomura:2010gx}
Nomura M, Kumagai N, Iwamoto S, Ota Y and Arakawa Y 2010 {\em Nat. Phys.\/}
  {\bf 6} 279

\bibitem{Alexander:2019um}
Alexander B and Christoph B 2019 {\em Nanophotonics\/} {\bf 8} 2041

\bibitem{ASTAKHOV2018211}
Astakhov G~V and Dyakonov V 2018 Defects for quantum information processing in
  sic {\em Defects in Advanced Electronic Materials and Novel Low Dimensional
  Structures\/} ed Stehr J, Buyanova I and Chen W (Woodhead Publishing) p 211

\bibitem{C7NR08249E}
Nguyen M, Kim S, Tran T~T, Xu Z~Q, Kianinia M, Toth M and Aharonovich I 2018
  {\em Nanoscale\/} {\bf 10}(5) 2267

\bibitem{SiV:APL:2011}
Neu E, Arend C, Gross E, Guldner F, Hepp C, Steinmetz D, Zscherpel E, Ghodbane
  S, Sternschulte H, Steinmüller-Nethl D, Liang Y, Krueger A and Becher C 2011
  {\em Appl. Phys. Lett.\/} {\bf 98} 243107

\bibitem{Lindner_2018}
Lindner S, Bommer A, Muzha A, Krueger A, Gines L, Mandal S, Williams O, Londero
  E, Gali A and Becher C 2018 {\em New J. Phys\/} {\bf 20} 115002

\bibitem{Mountford:08}
Mountford J, Porrovecchio G, Smid M and Smid R 2008 {\em Appl. Opt.\/} {\bf 47}
  5821

\bibitem{Park:16}
Park S, Hong K~S and Kim W~S 2016 {\em Appl. Opt.\/} {\bf 55} 2285

\bibitem{Bowman:14}
Bowman S~R, Brown C~G, Brindza M, Beadie G, Hite J~K, Freitas J~A, Eddy C~R,
  Meyer J~R and Vurgaftman I 2014 {\em Opt. Mater. Express\/} {\bf 4} 1287

\end{thebibliography}
\providecommand{\newblock}{}

\end{document}